\documentclass[prd,twocolumn,nofootinbib,notitlepage,aps,tightenlines,preprintnumbers,amsmath,amssymb,amsfonts,showpacs,superscriptaddress]{revtex4-2}

\RequirePackage[colorlinks=true
,urlcolor=blue
,anchorcolor=blue
,citecolor=blue
,filecolor=blue
,linkcolor=blue
,menucolor=blue
,linktocpage=true
,pdfproducer=medialab
,pdfa=true
]{hyperref}

\usepackage{amsmath,amssymb,amsthm,amsfonts}
\usepackage{graphicx,tabularx}
\usepackage{float}
\usepackage{multirow}  
\usepackage{cancel}
\usepackage{comment}
\usepackage{physics}
\usepackage{array}
\usepackage{enumitem}
\usepackage{soul}
\newcolumntype{P}[1]{>{\centering\arraybackslash}p{#1}}

\usepackage{cleveref}
\crefname{section}{Sec.}{Secs.}
\crefname{figure}{Fig.}{Figs.}
\crefname{equation}{Eq.}{Eqs.}
\crefname{table}{Table}{Tables}

\setlist[description]{leftmargin=0.3cm}
\setlist[itemize]{leftmargin=0.5cm}

\newcommand{\be}{\begin{equation} \begin{aligned}}
\newcommand{\ee}{\end{aligned} \end{equation}}

\newcommand{\mev}{\text{MeV}}
\newcommand{\gev}{\text{GeV}}
\newcommand{\tev}{\text{TeV}}

\newcommand{\cm}{\text{cm}}
\newcommand{\m}{\text{m}}

\usepackage{xcolor}
\RequirePackage[normalem]{ulem}

\begin{document}
\preprint{DESY-22-196, UCI-HEP-TR-2022-06, LA-UR-24-33427}

\title{\texorpdfstring{Neutrino Electromagnetic Properties and the Weak Mixing Angle \\ at the LHC Forward Physics Facility}{Neutrino Electromagnetic Properties and the Weak Mixing Angle at the LHC Forward Physics Facility}}

\author{Roshan Mammen Abraham}
\email{rmammen@okstate.edu}
\affiliation{Department of Physics,  Oklahoma State University, Stillwater, OK, 74078, USA}

\author{Saeid Foroughi-Abari}
\email{saeidf@uvic.ca}
\affiliation{Department of Physics and Astronomy, University of Victoria, Victoria, BC V8P 5C2, Canada}

\author{Felix Kling}
\email{felix.kling@desy.de}
\affiliation{Deutsches Elektronen-Synchrotron DESY, Germany}

\author{Yu-Dai Tsai}
\email{yt444@cornell.edu}
\email{ytsai@fnal.gov}
\affiliation{Los Alamos National Laboratory (LANL), Los Alamos, NM 87545, USA}

\begin{abstract}
The LHC produces an intense beam of highly energetic neutrinos of all three flavors in the forward direction, and the Forward Physics Facility (FPF) has been proposed to house a suite of experiments taking advantage of this opportunity.  In this study, we investigate the FPF's potential to probe the neutrino electromagnetic properties, including neutrino millicharge, magnetic moment, and charge radius. We find that, due to the large flux of tau neutrinos at the LHC, the FPF detectors will be able to provide more sensitive constraints on the tau neutrino magnetic moment and millicharge than previous measurements at DONUT, by searching for excess in low recoil energy electron scattering events.
We also find that, by precisely measuring the rate of neutral current deep inelastic scattering events, the FPF detectors have the potential to obtain the strongest experimental bounds on the neutrino charge radius for the electron neutrino, and one of the leading bounds for the muon neutrino flavor. The same signature could also be used to measure the weak mixing angle, and we estimate that $\sin^2 \theta_W$ could be measured to about $3\%$ precision at a scale $Q \sim 10~\gev$, shedding new light on the longstanding NuTeV anomaly.
\end{abstract}

\maketitle

\section{Introduction}

Neutrino properties are crucial to understanding our Universe and have been prime targets of particle physics experiments. The electromagnetic (EM) properties of neutrinos, in particular, can be tested in existing and future experiments.  These measurements include the mass-dimension 4 neutrino millicharge, the mass-dimension 5 neutrino dipole moments, and the mass-dimension 6 neutrino charge radius.  These properties can, for example, be used to determine whether neutrinos have a Dirac or Majorana nature~\cite{Shrock:1982sc, Frere:2015pma} and to probe new physics beyond the Standard Model (SM)~\cite{Giunti:2014ixa}. These neutrino properties could be linked to intriguing experimental anomalies, including the NuTeV anomaly~\cite{NuTeV:2001whx} and the Xenon 1T excess~\cite{Aprile:2020tmw} (although the latter was determined most likely to be from an SM background~\cite{XENON:2022ltv}). Large neutrino dipole moments, for example, can also affect the mass gap of black holes~\cite{Croon:2020ehi, Sakstein:2020axg}. Interesting models were proposed to generate neutrino EM couplings much larger than the SM predictions~\cite{Voloshin:1987qy, Barbieri:1988fh, Rajpoot:1990hj, Aboubrahim:2013yfa, Lindner:2017uvt, Babu:2020ivd} and to connect the anomalies to the neutrino properties~\cite{Babu:2021jnu}. Currently, the SM predictions of these properties are several orders of magnitude smaller than the present upper bounds, obtained from reactor neutrinos~\cite{Beda:2012zz,TEXONO:2006xds}, accelerator neutrinos~\cite{LSND:2001akn, DONUT:2001zvi, Allen:1992qe}, and solar neutrinos~\cite{Borexino:2008dzn, Borexino:2017fbd, Coloma:2022umy, XENON:2022ltv, A:2022acy, LZ:2022lsv}, to name a few. For a connection between neutrino electromagnetic properties and CP phases, see Ref.~\cite{AristizabalSierra:2021fuc}.

The LHC provides one of the most exciting opportunities in studying high-energy neutrinos and tau neutrinos, given its high center-of-mass energy. The forward region at the LHC, in particular, provides a large flux of neutrinos coming from meson decays~\cite{FASER:2019dxq}. The Forward Physics Facility (FPF)~\cite{Anchordoqui:2021ghd} at the LHC is ideally placed on studying these \tev~energy neutrinos. Previously, interesting signatures from the neutrino dipole portal~\cite{Magill:2018jla, Shoemaker:2020kji}, were studied at FPF~\cite{Ismail:2021dyp} and FASER~\cite{Jodlowski:2020vhr}, but a proper analysis of the future capability of FPF on neutrino EM properties are sorely lacking at this moment. 

In this letter, we utilize the FPF to study interesting properties of neutrinos: the neutrino millicharge, magnetic moment, and charge radius. By looking at low recoil energy electron scattering and neutral current deep inelastic scattering (DIS) events, we show that we can reach competitive sensitivity for these properties. Most excitingly, we can set the world's leading limit on neutrino charge radius for the electron neutrino, while for the muon neutrino, we come within a factor of a few from the SM prediction. For the tau neutrino, FPF's limits on the magnetic moment are an order of magnitude better than the DONUT results~\cite{DONUT:2001zvi} and bounds on millicharge and charge radius constitute some of the few existing measurements for the tau neutrino. 

The neutrino interactions with the target material, investigated in this study, also depend sensitively on electroweak parameters. In this context, the precise measurement of the neutral current neutrino DIS rate can also be translated to a precise measurement of the weak mixing angle. This would allow one to test the anomalous result obtained by NuTeV~\cite{NuTeV:2001whx}. 

The paper is organized as follows. We briefly review neutrino EM properties in \cref{nuEMprops} and introduce the detectors under consideration at the FPF in \cref{detectors}. In \cref{signal}, we discuss our signal characteristics. We present our results on the neutrino EM properties in \cref{results} and discuss the measurement of the weak mixing angle in \cref{WeakMixingAngle}. We conclude in \cref{conclusion}. 

\section{Neutrino EM Properties}\label{nuEMprops}

The electric charge of neutrinos is zero in the SM. However, electromagnetic properties can arise at the quantum loop level (or via BSM physics) allowing electromagnetic interactions of neutrinos with photons and charged particles. Considering neutrinos as massive fermions, the electromagnetic properties of neutrinos in the one-photon approximation can be assembled in the matrix element of the neutrino effective electromagnetic current~\cite{Kayser:1982br,Nieves:1981zt} as
\begin{equation}\label{EMCurrent}
\matrixel{\nu_f(p_f)}{j^{\mu}_{\nu,\rm{EM}}}{\nu_i(p_i)} = \overline{u}_f(p_f) \Lambda_{fi}^{\mu}(q) u_i(p_i),
\end{equation}
where $q$ is the four-momentum transferred to the photon. The vertex function $\Lambda_{fi}^{\mu}(q)$ is a $3\times 3$ matrix in the neutrino mass eigenstates space that encodes the electromagnetic properties of neutrinos. We are interested in the ultra-relativistic limit where, at low-$q^2$, it simplifies to,
\begin{equation}\label{VrtxSimp}
\Lambda_{fi}^{\mu}(q) = \gamma^{\mu}(Q_{fi}-\frac{q^2}{6}\expval{r^2}_{fi}) -i\sigma^{\mu\nu}q_\nu \mu_{fi}
\end{equation}
with $f=i$ for diagonal and $f~{\ne}~i$ for transition electromagnetic properties. Note that in theories of massive neutrinos, the transition electromagnetic properties can be generated through mixing, even if the matrices in \cref{VrtxSimp} are diagonal in the mass basis~\cite{Kouzakov:2017hbc}.

In this paper, we conduct a phenomenological study of effective neutrino electromagnetic properties: the millicharge $Q$, the magnetic moment $\mu$, and the charge radius $\langle r^2 \rangle$ at the FPF. Effective here implies the possible inclusion of contributions coming from electric and anapole moments to the magnetic moment and charge radius~\cite{Giunti:2022aea}, respectively. Also, the neutral current interaction we study here has no information on the outgoing neutrino flavor. Therefore, we implicitly assume a sum over all final state neutrino flavors~\cite{Grimus:2000tq,Beacom:1999wx}. Note that when recasting the results obtained here, e.g., for Majorana neutrinos, they have only transition magnetic moment and millicharge.
\medskip

The electric neutrality of neutrinos in the SM is guaranteed by charge quantization~\cite{Geng:1989tcu, Babu:1989ex}. But in some BSM theories, neutrinos can have a very small electric charge~\cite{Das:2020egb} enabling it to couple to the photon. This BSM interaction can be described by an effective term in the Lagrangian $\mathcal{L} \supset Q_{\nu} (\bar{\nu} \gamma_{\mu} \nu) A^{\mu}$. 
 
Neutrino magnetic moments, on the other hand, do arise at one loop level~\cite{Fujikawa:1980yx, Vogel:1989iv} for a massive neutrino. The diagonal magnetic moment for a massive Dirac neutrino is given by %
\begin{align}
\label{eq:nu_limit}
 \mu_{\nu}
 \approx \frac{3e G_F}{8 \sqrt{2} \pi^2} m_{\nu}
 \approx 3\cdot 10^{-19} \mu_B \left(\frac{m_{\nu}}{1{\rm ~eV}}\right).
\end{align}
where $m_\nu$ is the neutrino mass, $e$ is the electric charge, $G_F$ is the Fermi constant and $\mu_B = e/(2m_e)$ is the Bohr magneton. This very small value is beyond the scope of terrestrial and astrophysical probes currently. The values for transition magnetic moments for Majorana neutrinos are even smaller~\cite{Pal:1981rm}. However, an additional contribution to the magnetic moment of neutrinos could arise from BSM physics~\cite{Aboubrahim:2013yfa,Barbieri:1988fh,Mohapatra:2004ce}. In an effective field theory approach, this can be parametrized in terms of a higher dimensional operator $\mathcal{L} \supset \mu_{\nu} (\bar{\nu} \sigma_{\alpha\beta} \nu) F^{\alpha\beta}$ for Dirac neutrinos (for Majorana neutrinos one replaces $\bar{\nu}$ with $\bar{\nu}^{c}$ for only the left-handed neutrino fields ($\nu_L$) above, and only transition moments are allowed). 

Measuring the magnetic moment of neutrinos is important, as it can also in principle shed light on the Dirac vs. Majorana nature of neutrinos. Dirac neutrinos can have diagonal and transition magnetic moments, whereas Majorana neutrinos only have transition magnetic moments. Large transition magnetic moments for Majorana neutrinos could be realized in certain BSM models~\cite{Babu:1989wn,Barr:1990um}, which are not too far from the current experimental limits, but the off-diagonal moments could be hard to measure, as we do not probe the outgoing neutrino flavor. 

Neutrinos also have non-zero charge radii in the SM from radiative corrections given by~\cite{Bernabeu:2002pd, Bernabeu:2000hf} 
\begin{equation}
\label{eq:nu_charge_radii}
\left<r^2_{\nu_{\ell}}\right>_{\rm SM}= \frac{G_F}{4\sqrt{2}\pi^2}\left[3-2 \log\frac{m_{\ell}^2}{m_W^2}\right].
\end{equation}
where $m_{\ell}$ are the lepton masses ($\ell=e,\mu,\tau$) and $m_W$ is the W boson mass. The SM values are then found to be $4.1 \times 10^{-33}~{\rm cm}^2$ for $\nu_e$, $2.4\times 10^{-33}~{\rm cm}^2$ for $\nu_\mu$ and $1.5 \times 10^{-33}~{\rm cm}^2$ for $\nu_\tau$. These values differ by at most one or two orders of magnitude from current terrestrial bounds, and hence testing the SM prediction of neutrino charge radius is a compelling challenge.

\section{Detectors at the FPF}\label{detectors}

An unexpected but powerful source of light and weakly coupled particles can be found at the LHC~\cite{Feng:2017uoz}. In the forward direction, the LHC produces an intense and strongly collimated beam of neutrinos of all three flavors coming mainly from the decays of mesons produced at the interaction point. Currently, there are two experiments taking advantage of this opportunity: FASER$\nu$~\cite{FASER:2019dxq, FASER:2020gpr} and SND@LHC~\cite{Ahdida:2750060, SHiP:2020sos}. In particular, both experiments are expected to obtain about 20 tau neutrino interactions, which exceeds the number of events recorded by the DONuT~\cite{DONuT:2007bsg} and OPERA~\cite{OPERA:2018nar} experiments. 

Several improved neutrino detectors are planned for the HL-LHC era. They will be housed in the FPF~\cite{Anchordoqui:2021ghd, Feng:2022inv} along with an array of other detectors with a wide range of physics potential, to be located in a cavern 620 m downstream from the ATLAS interaction point. Our analysis focuses on two detector technologies at FPF that are sensitive to TeV range neutrino interactions: FLArE, a liquid argon time projection chamber, and FASER$\nu$2, an emulsion-based neutrino detector. In the following, we present the detector details relevant to the phenomenological study at hand: 

\begin{itemize}[leftmargin=0.4cm]
\item \textbf{FLArE}, the Forward Liquid Argon Experiment, is composed of a 10-tonne liquid argon time projection chamber with a fiducial volume of $1\m \times 1\m \times 7\m$~\cite{Batell:2021blf}. Liquid argon time projection chambers are a proven technology for neutrino physics, having been used at Fermilab's Short-Baseline Neutrino Program~\cite{MicroBooNE:2015bmn} and at the future DUNE experiment~\cite{DUNE:2020ypp}. They offer the dual advantage of very low energy thresholds of down to $30~\mev$ and excellent timing resolution, achieved through a light collection system. This will allow one to control possible muon-induced backgrounds by vetoing events in coincidence with a muon track, which is critical to the feasibility of our study. We also include in our study a larger 100-tonne detector, dubbed \textbf{FLArE-100}, with a fiducial volume of $1.6\m \times 1.6\m \times 30\m$. This is meant to illustrate how sensitivities would scale with target mass.
\item \textbf{FASER$\nu$2} is an emulsion detector designed as a much larger successor to the approved FASER$\nu$ detector~\cite{SnowmassFASERnu2}. In the HL-LHC era, FASER$\nu$2 is envisioned as a 10-tonne neutrino detector composed of emulsion layers interleaved with tungsten sheets acting as target material. Emulsion detectors are capable of detecting charged tracks with high spatial resolution. The major drawback of emulsion detectors is a lack of timing information associated with the recorded events. FASER$\nu$2 aims to mitigate this by introducing tracking layers between and at the end of the emulsion layers. Timing information can then be obtained by successfully matching the event in the emulsion and the tracker. This is helpful in the search for a coincident muon track, which can be used to reduce muon-induced backgrounds. We assume that in FASER$\nu$2 all muon-induced backgrounds can be eliminated with the help of timing information. The fiducial volume we consider is $0.5~\m \times 0.5~\m \times 2~\m$~\cite{Batell:2021aja}. Since the charged particle has to pass through a sufficient number of emulsion layers to leave a distinguishable track, a minimum particle momentum of $300~\mev$~\cite{Batell:2021aja} is required. This sets the energy threshold of the detector.
\end{itemize}

The two processes we study here are neutrino electron elastic scattering and neutral current DIS. The main backgrounds for the former are similar to those studied in Refs.~\cite{Batell:2021blf, Batell:2021snh} and the latter was studied in the context of FASER$\nu$ in Ref.~\cite{Ismail:2020yqc}. Here we briefly summarize the relevant results. 

A major source of similar backgrounds for both processes is muon-induced events. Muons passing through the detector can, for example, emit photons through bremsstrahlung or produce high energy neutral hadrons in inelastic scatterings. The photons could then pair convert to $e^+e^-$ and if one of them is missed, it can mimic our electron scattering signal. Neutral hadron scattering, on the other hand, would look similar to the neutral current DIS neutrino interactions. In both cases, the inclusion of timing capabilities in the detectors allows vetoing such backgrounds by associating such events with the accompanying muon. For example, the currently operating FASER detector employs several scintillating veto layers at its front, each of which has a muon detection efficiency of more than $99.99\%$~\cite{FASER:2022hcn}. In this paper, we assume such muon-induced backgrounds can be reduced to negligible levels. 

An irreducible source of backgrounds to both processes is SM contribution to neutral current neutrino scattering. For neutrino electron elastic scattering, we employ the use of kinematic cuts to enhance the signal to background ratio, as described in the next section. This is where a low energy threshold detector like FLArE is advantageous.
\medskip

Throughout this work, we use the neutrino fluxes presented in Ref.~\cite{Batell:2021aja} for the HL-LHC era. They were obtained using the event generator \texttt{SIBYLL~2.3d}~\cite{Ahn:2009wx, Riehn:2015oba, Riehn:2017mfm, Riehn:2019jet} implemented via \texttt{CRMC}~\cite{CRMC} which simulates the primary collision. Ref.~\cite{Kling:2021gos} introduced a fast neutrino flux simulator that models the propagation and decay of long-lived hadrons within the SM in the forward direction at the LHC. Currently, there exist sizeable uncertainties on the neutrino flux. However, this is expected to be brought under control using the charged current scattering event rate once the detector starts to take data~\cite{FASER:2019dxq}.

\section{Neutrino EM Interaction Rate}\label{signal}

The signature we investigate in our study is the excess (or deficit) of neutrino scattering events in the detectors compared to the expected rate predicted by the SM in the absence of any neutrino EM properties. This can be neutrino electron elastic scattering events or neutral current DIS events.
\medskip 

We first consider the neutrino electron elastic scattering where the SM cross section, in terms of the electron recoil energy $E_r$, is given by~\cite{Formaggio:2012cpf, Giunti:2014ixa} 
\begin{equation}
\begin{aligned}\label{CrossSectionSM}
\bigg(\!\frac{d\sigma_{\nu_{\ell} e}}{dE_{r}}\!\bigg)_{\rm \!\!SM} \!\!\!\!\!\ = & \ \frac{ G_F^2 m_e}{ 2\pi} \bigg[  (g^{\ell}_{V} \!-\! g^{\ell}_{A})^2 \left(\! 1{-}\frac{E_{r}}{E_\nu}\right)^{2} 
\\
&+ (g^{\ell}_{V} \!+\! g^{\ell}_{A})^2 + \big((g^{\ell}_{A})^2 {-} (g^{\ell}_{V})^2 \big) \frac{m_e E_{r}}{E_\nu^2}\bigg] 
\end{aligned}
\end{equation}
with the standard vector and axial vector coupling constants $g_A^\ell$ and $g_A^\ell$ given by 
\begin{align}\label{VAcouplingconstants}
g^{\ell}_{V} = 2\sin^2 \theta_W - \frac{1}{2} + \delta_{\ell e} \text{ , } g^{\ell}_{A} = -\frac{1}{2} + \delta_{\ell e} \ .
\end{align}
Here $G_F$ is the Fermi constant, $\theta_W$ is the weak mixing angle, and $E_\nu$ is the neutrino energy. For antineutrinos, one must replace $g_A^\ell$ by $-g_A^\ell$. There is an extra term for the electron neutrino coming from the exchange of the $W$ boson, which is not present for muon and tau neutrinos. In the presence of non-negligible values for the neutrino electromagnetic properties, the event rate and distribution can be sufficiently distorted. 

As detailed below, the most significant effect of including these BSM physics is in the event rate, especially at low recoil energies for the magnetic moment and millicharge. This motivates looking at $E_r$ as the main kinematic variable in our study. One could also look at the recoil angle of the electron, as was studied in Ref.~\cite{Batell:2021blf}. For neutrino electron scatterings at the energies of interest, so $E_{\nu}$ and $ E_r \gg m_e$, the recoil angle is correlated with the recoil energy via $\cos\theta_r \approx 1-m_e/E_r$. Although, this does not help to distinguish different neutrino electron scattering events but provides another handle to remove backgrounds coming from neutrino nuclear scattering events with a single particle recoiling in the final state. Since we will be imposing a strong kinematic cut on the electron recoil energy that suppresses the background sufficiently, we do not include the recoil angle of the electron as an additional observable. We note, however, that the strong correlation between the recoil energy and the recoil angle of the electron can be used to improve energy resolution at small energies. 
\medskip 

In some cases, it might be beneficial to also consider nuclear scattering, where one could benefit from higher event rates. 
As can be seen in \cref{CrossSectionSM}, the neutrino electron elastic scattering cross section and hence the event rate is proportional to the target mass $m_e$.
If the new physics signal count decreases or does not increase commensurately, then moving to a heavier target will only degrade the sensitivity. This is the case with neutrino magnetic moment and millicharge, and hence we stick to electron scattering events for both of them.  As we will see below, a charge radius essentially induces a shift in the vector coupling constant, $g_V$, and hence we can expect higher rates of signal if we use a heavier target. We therefore also consider neutral current neutrino DIS in the charge radius case, which will result in significantly higher signal event rates and hence improve the bounds on $\langle r_{\nu_{\ell} }^2\rangle$.

At leading order, the double differential cross section for neutral current neutrino-nucleon DIS is given by~\cite{McFarland:2008xd}
\begin{equation}
\begin{aligned}
\label{nuDIS}
&\frac{d\sigma(\nu N \to \nu X)}{dx \ dy} = \frac{ 2G_F^2 m_p E_\nu}{ \pi} \frac{m_{Z}^4}{(Q^2+m_{Z}^2)^2} \times
\\
&\sum_{q=u,d,s,c} \big[ g_{q,L}^2 [x f_{q}(x,Q^2) + x f_{\bar{q}}(x,Q^2)(1-y)^2 ] +
\\
& \ \ \ \ \ \ \ \ \ \ \ \   g_{q,R}^2 [x f_{q}(x,Q^2) (1-y)^2 + x f_{\bar{q}}(x,Q^2) ] \big ]
\end{aligned}
\end{equation}
for neutrino scattering and
\begin{equation}
\begin{aligned}
\label{nubarDIS}
&\frac{d\sigma(\overline{\nu} N \to \overline{\nu} X)}{dx \ dy} = \frac{ 2G_F^2 m_p E_\nu}{ \pi} \frac{m_{Z}^4}{(Q^2+m_{Z}^2)^2} \times
\\
&\sum_{q=u,d,s,c} \big[ g_{q,L}^2 [x f_{q}(x,Q^2)(1-y)^2 + x f_{\bar{q}}(x,Q^2)] +
\\
& \ \ \ \ \ \ \ \ \ \ \ \  g_{q,R}^2 [x f_{q}(x,Q^2) + x f_{\bar{q}}(x,Q^2) (1-y)^2] \big ]
\end{aligned}
\end{equation}
for anti-neutrino scattering, where $X$ stands for the final states that are a byproduct of the DIS other than the neutrino. Here $m_p$ is the mass of the target proton, $m_Z$ is the $Z$ boson mass, and $g^q_{L}, g^q_{R} = T^3 - Q_q \sin^2 \theta_W$ are the left and right-handed neutral current couplings of the quarks with $Q_q$ being the charge of the quarks in units of $e$, and $T^3$ being the third component of the quarks's weak isospin. The differential cross section is expressed in terms of the DIS variables $x$, $y$ and $Q^2$, where $x$ is the partonic momentum fraction, $y = E_\text{had}/E_\nu$ is the fraction of neutrino's energy that is transferred to the hadronic system, and $Q^2 = 2 m_p E_\nu x y$ is the squared 4-momentum transfer. Here $E_{\nu}$ is the incident neutrino energy, and $E_{\rm{had}}$ is all the energy contained in the hadronic system. The functions $f_q(x,Q^2)$ are the nucleon parton distribution function. Here we use nCTEQ15 which includes nuclear effects of the target nucleus~\cite{Kovarik:2015cma}. 
\begin{figure*}[phtb]
    \includegraphics[width=0.49\textwidth]{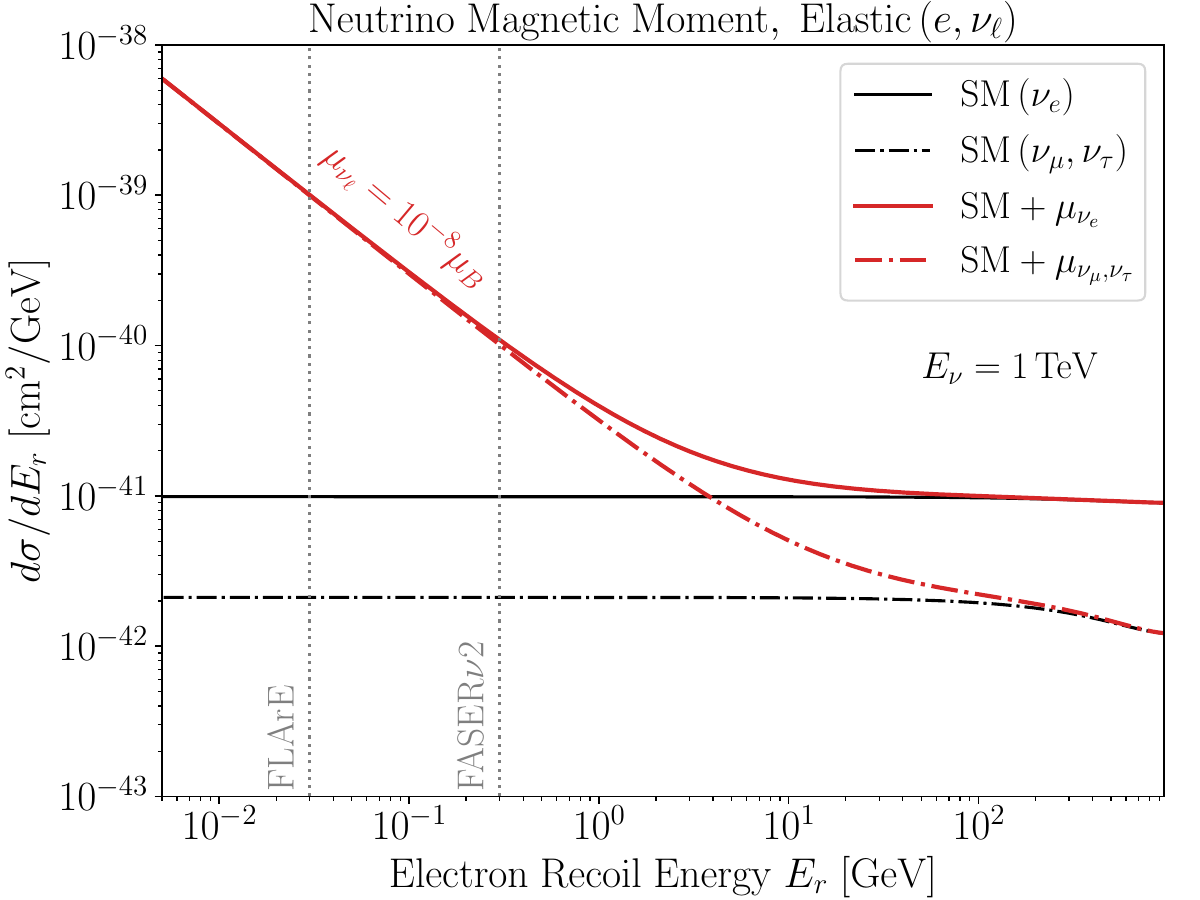}
    \includegraphics[width=0.49\textwidth]{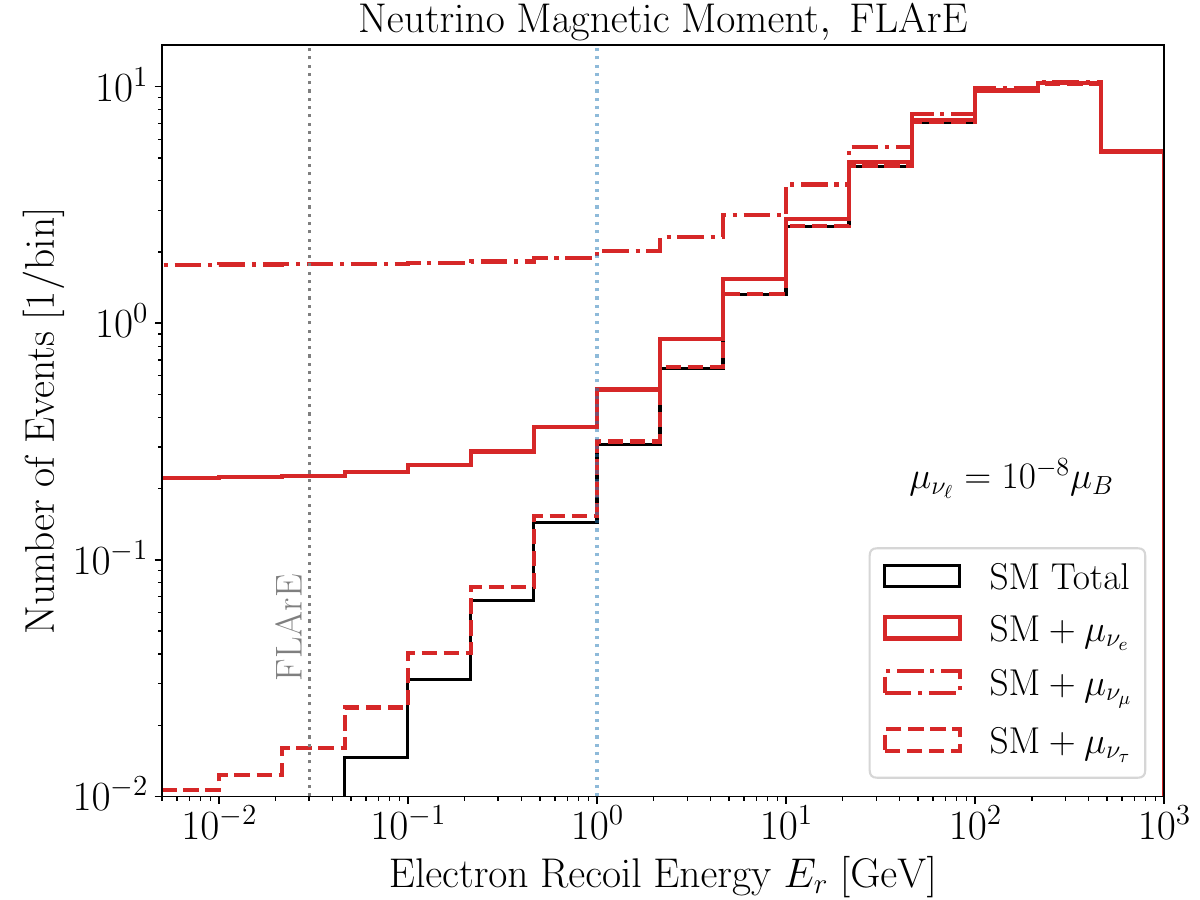} 
    \\
    \vspace{0.1cm}
    \includegraphics[width=0.49\textwidth]{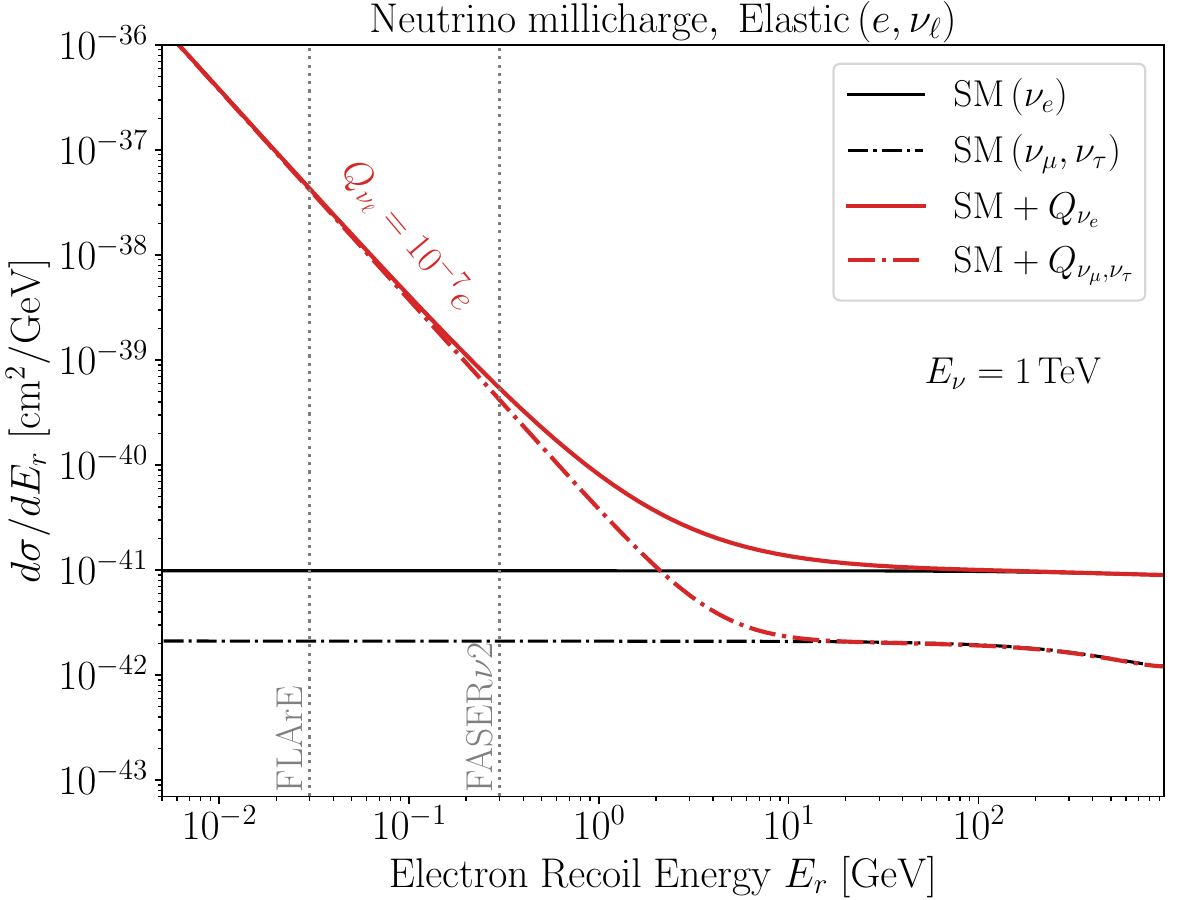}
    \includegraphics[width=0.49\textwidth]{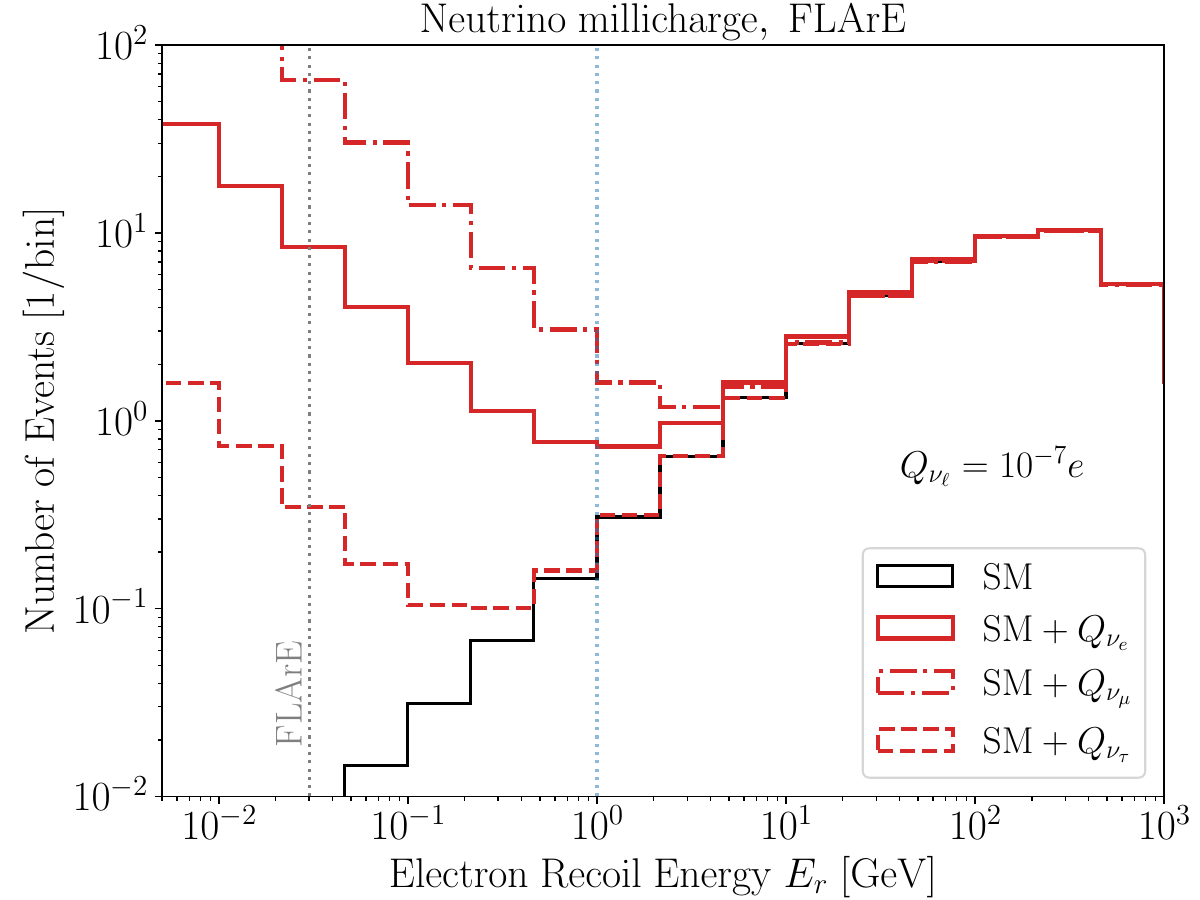}
    \\
    \vspace{0.1cm}
    \includegraphics[width=0.49\textwidth]{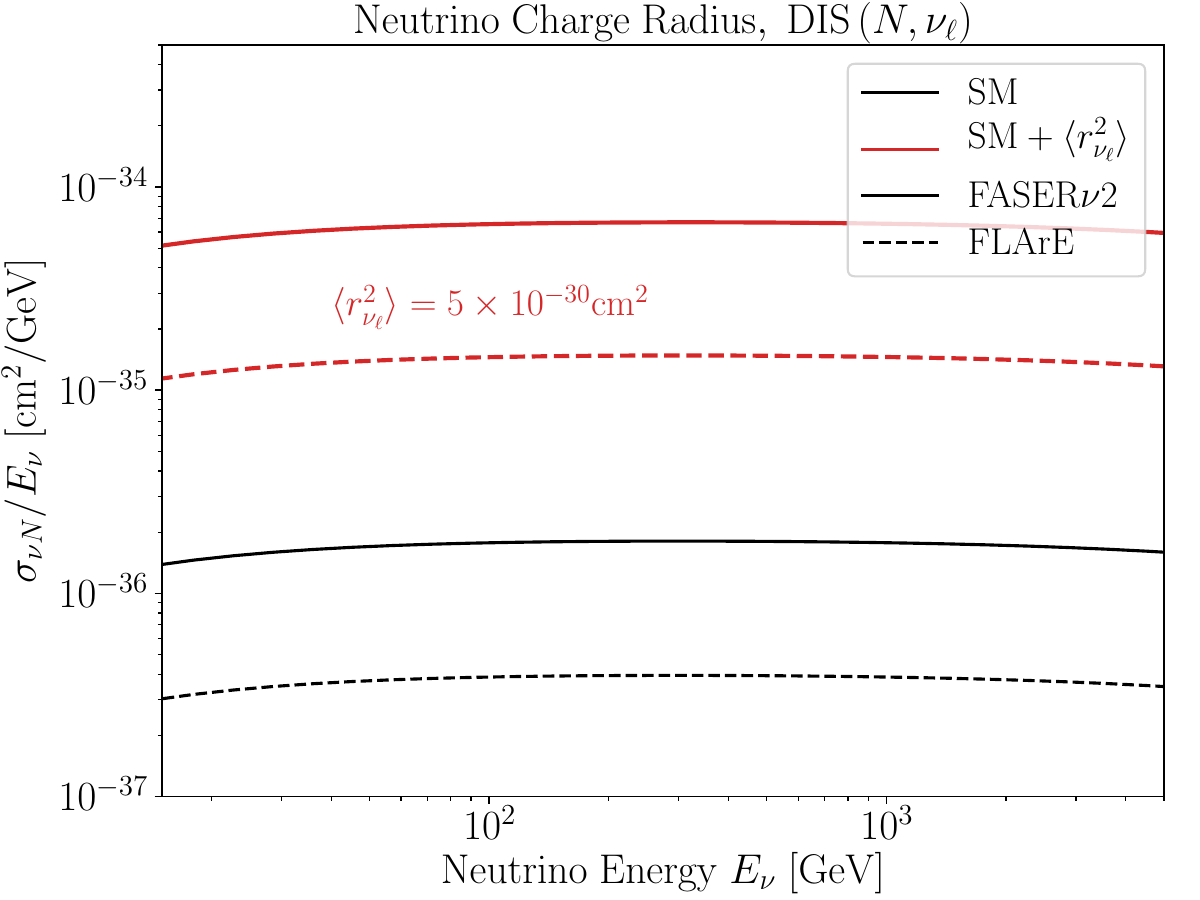}
    \includegraphics[width=0.49\textwidth]{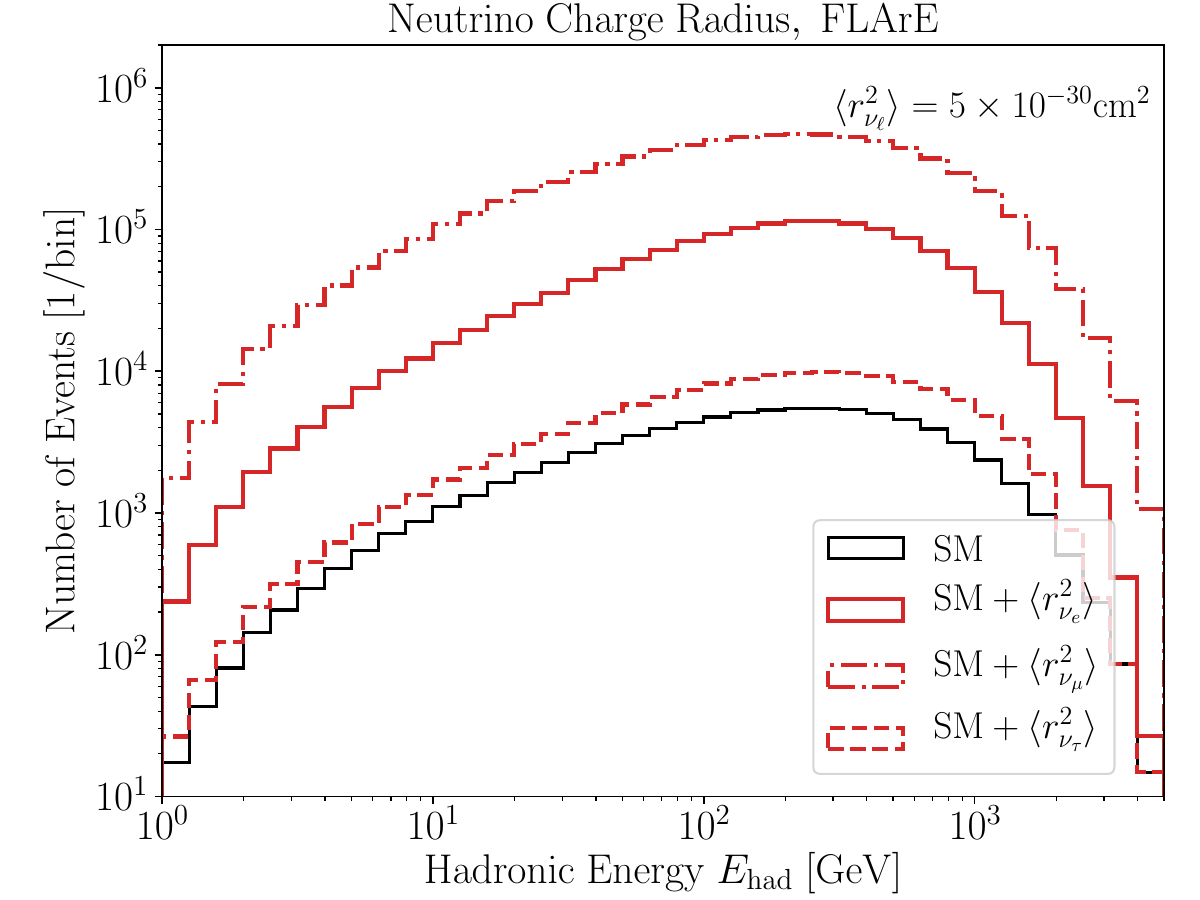} 
    \caption{\textbf{Top and Middle Left:} Differential cross-section of neutrino elastic scattering on the electron as a function of electron recoil energy, corresponding to the incoming neutrino energy of 1~TeV. \textbf{Top and Middle Right:} The expected number of events at FLArE, considering the estimated neutrino flux at the FPF in the HL-LHC phase. The magnetic moment and millicharge electromagnetic contributions (red) exceed the SM background (black) at lower recoil energies. The FLArE and FASER$\nu$2 detector recoil energy thresholds of 30 and 300 $\mev$, as well as the $1~\gev$ upper cutoff, are indicated by vertical dotted lines. \textbf{Bottom Left:} Cross section of neutrino-nucleus deep-inelastic scattering in FASER$\nu$2 (solid) and FLArE (dashed) as a function of neutrino energy, within the SM (black) and in the presence of the charge radius. \textbf{Bottom Right:} Expected event rate at FLArE as a function of the energy of the hadronic system. 
    }
    \label{fig:rates}
\end{figure*}

\subsection{Neutrino Magnetic Moment}

The presence of a BSM contribution to the neutrino magnetic moment can lead to an excess in the number of electron recoil events, especially at low recoil energies. The differential cross section with respect to the electron recoil energy for the elastic scattering of a neutrino (or antineutrino) with incoming flavor $\ell$ and energy $E_{\nu}$ off an electron, in the presence of a magnetic moment, is given by~\cite{deGouvea:2006hfo,Babu:2020ivd}
\begin{align}
\label{CrossSectionNMM}
\!\!\!\!\ \bigg(\! \frac{d\sigma_{\nu_{\ell} e}}{dE_{r}} \!\bigg)_{\rm \!\!NMM}\!\!\!\!\!  = \bigg(\! \frac{d\sigma_{\nu_{\ell} e}}{dE_{r}}\!\bigg)_{\rm \!\! SM} \!\!\!\! + \frac{\pi  ^2}{m_e ^2}  \bigg(\!\frac{1}{E_{r}} {-} \frac{1}{E_\nu}\!\bigg)\!\bigg(\!\frac{\mu_{\nu_{\ell}}}{\mu_{\rm B}}\!\bigg)^2 \!\! ,
\end{align}
where $\mu_{\nu_{\ell}}$ is the effective neutrino magnetic moment, and $\mu_B$ is the Bohr magneton. Note that the two contributions in \cref{CrossSectionNMM} add incoherently in the cross section due to the following helicity argument~\cite{Grimus:1997aa}: in the ultra-relativistic limit, the SM weak interaction conserves the neutrino helicity while the helicity flips in the neutrino magnetic moment interaction. Hence, one is always guaranteed an excess of events in this case.

The two contributions in the cross section exhibit quite different dependencies in the electron recoil energy $E_r$, as illustrated in \cref{fig:rates} top left panel for an incoming neutrino beam with $1~\tev$ energy. The signal cross section associated with the neutrino magnetic moment exceeds the SM background in the range, 
\begin{align}
    E_r &\lesssim 
    10~{\rm GeV} \times  \bigg(\frac{\mu_{\nu_{\ell}}}{10^{-8} \, \mu_{\rm B}}\bigg)^2.
\end{align}
This leads to an increase in the elastic neutrino-electron events above the SM predicted value at low values of $E_r$. This can be seen as arising from the $1/E_r$ term in the BSM cross section expression. The lines for $\nu_e$ and $\nu_{\mu,\tau}$ differ due to the additional diagram from the W boson exchange, which is only present for $\nu_e$.

Given the neutrino flux at the FPF, the electron recoil energy spectrum at FLArE and FASER$\nu$2 detectors can be calculated. \cref{fig:rates} top right panel shows the total expected event rate for a benchmark value of $\mu_{\nu_{\ell}}=10^{-8}~\mu_B$ for all three flavors at FLArE, as well as the SM event rate. The three flavors of neutrinos have different fluxes at FPF, resulting in distinct predictions for the event spectrum.
The excess events in the low recoil energy bins serve as an experimental signature to look for neutrino magnetic moment. 

\subsection{Neutrino Millicharge}

The FPF is an ideal environment to search for millicharged particles in the dark sector~\cite{Foroughi-Abari:2020qar, Kling:2022ykt} and can also be used to probe neutrino millicharge. The inclusion of a non-zero electric charge for the neutrino changes the neutrino-electron elastic scattering cross section as~\cite{Kouzakov:2017hbc, Mathur:2021trm, Giunti:2014ixa}
\begin{align}\label{CrossSectionNMC}
\!\!\!\!\!\bigg( \!\frac{d\sigma_{\nu_{\ell} e}}{dE_{r}} \!\bigg )_{\rm \!\!NMC} \!\!\!\!\!=  \ \bigg( \! \frac{d\sigma_{\nu_{\ell} e}}{dE_{r}} \!\bigg )_{\rm \!\!SM} \!\!\!\!\! + \bigg (\!\frac{d\sigma_{\nu_{\ell} e}}{dE_{r}} \!\bigg )_{\rm \!\!Int} \!\!\!\!\! + \bigg (\!\frac{d\sigma_{\nu_{\ell} e}}{dE_{r}} \!\bigg )_{\rm \!\!Quad} \!\!\!\!\! . 
\end{align}
The first term is the above SM expression as before. The interference term is,
\begin{align}
\bigg (\!\frac{d\sigma_{\nu_{\ell} e}}{dE_{r}} & \!\bigg )_{\rm Int}\!\!\!\!\! = \ \frac{\sqrt{8\pi}G_F \alpha}{E^{2}_{\nu}E_{r}}\bigg(\frac{Q_{\nu_{\ell}}}{e}\bigg)\bigg[ g^{\ell}_{V} \big(2E^{2}_\nu + E^{2}_{r} 
\nonumber
\\
&- E_{r}(2E_\nu + E_{r})\big) + g^{\ell}_{A}\big(E_{r}(2E_\nu - E_{r})   \big)  \bigg]
\end{align}
with $g^{\ell}_{V}, g^{\ell}_{A}$ defined as before, and the quadratic term is given by 
\begin{equation}
\!\!\!\!\! \bigg (\!\frac{d\sigma_{\nu_{\ell} e}}{dE_{r}}\!\bigg )_{\rm \!\! Quad} \!\!\!\!\!\!\!\! = 4(\pi \alpha)^2 \bigg(\!\frac{Q_{\nu_{\ell}}}{e} \!\bigg)^{\!2} \bigg[ \frac{2E^{2}_{\nu} {+} E^{2}_{r} {-} 2E_{\nu}E_{r} }{m_e E^{2}_{r} E^{2}_{\nu} } \bigg],
\end{equation}
where $Q_{\nu_{\ell}}$ is the electric charge of the neutrino in units of $e$. For anti-neutrino, we replace $g_A$ with $-g_A$ similar to the SM case. The presence of the interference term $\sim Q_{\nu_{\ell}}$ means we are now sensitive to the sign of the neutrino millicharge and depending on the value of $Q_{\nu_{\ell}}$ we can expect an increase or decrease in the number of events. However, it turns out that, for the values of the millicharge that can be probed at the FPF, the quadratic term always dominates, therefore, an excess of events is expected. For a benchmark value of $Q_{\nu_{\ell}}=10^{-7}e$ we see an even steeper increase in cross section at lower recoil energies than for the magnetic moment, as the quadratic term grows proportionally to $1/E_r^2$. This is shown in \cref{fig:rates} middle left panel, while the middle right panel shows the event spectrum at FLArE. 

Alternatively, neutrino millicharge can also be probed at FORMOSA~\cite{Foroughi-Abari:2020qar}, a proposed experiment located within the FPF to search for millicharged particles. If neutrino possesses a millicharge, then it will ionize the material and deposit energy as it passes through the detector, resulting in a scintillation signature. FORMOSA is a dedicated detector to detect low-charge scintillation signals consisting of an array of plastic scintillators with multiple layers, sensitive to low-energy deposits down to one single photoelectron. The mean ionization energy loss for a millicharged neutrino travelling through the plastic material can be estimated by Bethe-Bloch formula ~\cite{Workman:2022ynf} that goes as $\langle dE/dx\rangle \sim (Q_\nu/e)^2\times 5 \,\mev/\cm$, and is quite insensitive to the neutrino energy and mass. The average number of photoelectrons produced within a scintillator bar $N_{\rm{PE}}$ is proportional to the ionization energy deposition, the bar length, and the scintillation light yield. To suppress the background noise in the photomultiplier tubes attached to the scintillation bars that collect the produced photoelectrons, the low-energy scintillation signal candidates are required to have multiple coincidences of hits. To detect a millicharged neutrino, at least one photoelectron in each layer of the scintillator must be observed. The detection probability $P=(1-\exp(-N_{\rm{PE}}))^n$ follows the Poisson distribution, where $n$ is the number of layers.

\subsection{Neutrino Charge Radius}

From \cref{VrtxSimp}, one sees that a non-zero value of charge radius amounts to a shift in the vector term of the neutrino vertex function. Within the SM, only diagonal charge radii are allowed, as generation lepton numbers are conserved. However, some BSM scenarios also allow for off-diagonal charge radii~\cite{Cadeddu:2020lky, Bernstein:1963jp, Grau:1985cn}. If we only consider diagonal elements in the flavor basis, it was shown in Ref.~\cite{Vogel:1989iv} that this amounts to a modification of the vector coupling constant in \cref{VAcouplingconstants} as,
\begin{equation}\label{VcouplingshisftfromCR}
g^{\ell}_{V}\rightarrow g^{\ell}_{V} + \frac{2}{3}m^2_W\langle r_{\nu_{\ell}}^2\rangle \sin^2 \theta_W.
\end{equation}
This introduces additional linear and quadratic terms in $\langle r_{\nu_{\ell}}^2\rangle$ to the cross section in \cref{CrossSectionSM}. Therefore, similar to the neutrino millicharge case, we are sensitive to the sign of $\langle r_{\nu_{\ell}}^2\rangle$. Also, note that the antineutrino charge radius contribution comes with a negative relative sign to the above shift~\cite{Giunti:2014ixa}.
For quarks, this shift is modified by the quark-to-electron electric charge ratio as,
\begin{equation}\label{VcouplingshiftfromWMAforquarks}
g^{q}_{V}\rightarrow g^{q}_{V} - \frac{2}{3}Q_q m^2_W\langle r_{\nu_{\ell}}^2\rangle \sin^2 \theta_W
\end{equation}
which modifies the left and right-handed neutral current couplings of the quarks $g^q_{L/R}=(g^q_V \pm g^q_A)/2$ in \cref{nuDIS,nubarDIS}. 

In the bottom left panel of \cref{fig:rates}, we show the DIS cross section rates for a neutrino scattering off the argon nucleus in the FLArE detector and tungsten nucleus in the FASER$\nu$2 detector as a function of the incoming neutrino energy, $E_\nu$. FASER$\nu$2 with a target atom with a higher atomic number has more nucleons for the neutrino to scatter off and hence has a higher cross section value. In the presence of a non-zero charge radius, the cross section enhancement is almost uniform across the incoming neutrino energy range. In the bottom right panel, we show the event spectrum as a function of $E_{\rm{had}}$ at FLArE for a benchmark value of $\langle r_{\nu_{\ell}}^2 \rangle=5\times 10^{-30}~{\rm cm}^2$. At the neutrino energies available at FPF and the values of $\langle r_{\nu_{\ell}}^2\rangle$ we are sensitive to, it is the quadratic term that is dominant, and we observe an excess in events across the spectrum.

\section{Sensitivity for Neutrino EM Properties}\label{results}

\begin{table*}[t!]
\begin{tabular}{ P{3cm}|P{3cm}||P{2.5cm}|P{2.5cm}|P{2.5cm} } 
\hline
\hline
\multicolumn{2}{c||}{ Neutrino EM Property } & FASER$\nu$2 & FLArE & FLArE-100 \\
 \hline
 \hline
\multirow{3}{*}{ $\mu_{\nu_{\ell}}$ $[10^{-8}\mu_B]$ } & $\nu_e$ & 1.78 & 1.35 & 0.73 \\ 
& $\nu_\mu$ & 0.67 & 0.48 & 0.25 \\ 
& $\nu_\tau$ & 10.7 & 6.59 & 3.08 \\ 
 \hline
\multirow{3}{*}{ $Q_{\nu_{\ell}}$ $[10^{-8}e]$ } & $\nu_e$ & [-13.1 , 8.92] & [-4.03 , 3.21] & [-2.21 , 1.52] \\ 
& $\nu_\mu$ & [-3.92 , 4.12] & [-0.96 , 1.27] & [-0.24 , 0.30] \\ 
& $\nu_\tau$ & [-64.9 , 65.1] & [-17.9 , 17.9] & [-8.33 , 8.36] \\ 
 \hline
 \multirow{2}{*}{ $\langle r^2_{\nu_{\ell}}\rangle$ $[10^{-32} \mathrm{cm}^2]$ }  & $\nu_e$& [-3.57 , 4.46] & [-3.47 , 4.29] & [-1.43 , 1.55] \\ 
& $\nu_\mu$ & [-0.65 , 0.67] & [-0.62 , 0.64] & [-0.25 , 0.25] \\ 
Nuclear Scattering & $\nu_\tau$ & [-58.9 , 96.1] & [-41.3 , 78.4] & [-17.3 , 54.8] \\ 
 \hline
\multirow{2}{*}{ $\langle r^2_{\nu_{\ell}}\rangle$ $[10^{-31} \mathrm{cm}^2]$ } & $\nu_e$ & [-1.11 , 0.85] & [-1.62 , 1.10] & [-0.54 , 0.47] \\ 
& $\nu_\mu$ & [-0.86 , 1.70] & [-1.03 , 1.79] & [-0.56 , 1.29] \\ 
Electron Scattering
& $\nu_\tau$ & [-16.4 , 16.6] & [-14.5 , 14.8] & [-7.53 , 8.04] \\ 
 \hline
 \hline
\end{tabular}
\caption{Projected 90\% C.L. sensitivity on neutrino electromagnetic properties  ($\mu_{\nu_{\ell}}$, $Q_{\nu_{\ell}}$, $\langle r^2_{\nu_{\ell}}\rangle$)  from FASER$\nu$2, FLArE, FLArE-100 detectors for all three flavors, assuming 3 ab$^{-1}$ of integrated luminosity at HL-LHC. For completeness, we also show the charge radius bounds from electron scattering in the last row, which, as expected, are much weaker compared to those from nuclear scattering.
}
\label{tab:results}
\end{table*}

\begin{table*}[t!]
\begin{tabular}{ P{3cm}|P{3cm}||P{2.5cm}|P{2.5cm}|P{2.5cm} } 
\hline
\hline
\multicolumn{2}{c||}{ Neutrino Charge Radius} & FASER$\nu$2 & FLArE & FLArE-100 \\
 \hline
 \hline

 \multirow{2}{*}{ $\langle r^2_{\nu_{e}}\rangle$ $[10^{-32} \mathrm{cm}^2]$ }  
& $3\%$ & [-13.5 , 50.1] & [-13.8 , 51.1] & [-13.5 , 51.4] \\ 
 & $1\%$ & [-6.05 , 9.16] & [-6.09 , 9.11] & [-5.35 , 7.33] \\ 
 & $0.3\%$ & [-3.86 , 4.92] & [-4.31 , 5.60] & [-3.04 , 3.58] \\ 
 & $0\%$ & [-3.57 , 4.46] & [-3.47 , 4.29] & [-1.43 , 1.55] \\  
 \hline

 \multirow{2}{*}{ $\langle r^2_{\nu_{\mu}}\rangle$ $[10^{-32} \mathrm{cm}^2]$ }
& $3\%$ & [-2.93 , 3.03] & [-2.93 , 3.04] & [-2.90 , 3.01] \\ 
 & $1\%$ & [-1.15 , 1.18] & [-1.14 , 1.17] & [-1.00 , 1.01] \\ 
 & $0.3\%$ & [-0.70 , 0.72] & [-0.78 , 0.80] & [-0.54 , 0.55] \\  
 & $0\%$ & [-0.65 , 0.67] & [-0.62 , 0.64] & [-0.25 , 0.25] \\  
 \hline

 \multirow{2}{*}{ $\langle r^2_{\nu_{\tau}}\rangle$ $[10^{-32} \mathrm{cm}^2]$ }
& $3\%$ & [-146.4 , 183.6] & [-109.9 , 146.9] & [-91.1 , 128.6] \\ 
 & $1\%$ & [ -84.0 , 121.2] & [ -61.3 , 98.4] & [-46.3 , 83.8] \\ 
 & $0.3\%$ & [-62.1 , 99.3] & [-48.1 , 85.2] & [-30.7 , 68.1] \\  
 & $0\%$ & [-58.9 , 96.1] & [-41.3 , 78.4] & [-17.3 , 54.8] \\  
 \hline 
 
 \hline
\end{tabular}
\caption{
Same as \cref{tab:results} but only for neutrino charge radius from nuclear scattering. Now we include an uncertainty of 3\%, 1\%, 0.3\%, and 0\% on the neutrino-nucleus cross section. Results for 0\% are the same as those shown in \cref{tab:results}.
}
\label{tab:NCR_systematics}
\end{table*}

We are now ready to turn to our analysis. As described in the previous section, both the neutrino magnetic moment and neutrino millicharge would manifest themselves through an enhanced rate of neutrino-electron scattering events with low electron recoil energy. To isolate this effect, we select events within the energy range $E_{\rm{thr}} < E_r < 1~\gev$. Here we assume a lower energy threshold of $E_{\rm{thr}}=30~\mev$ for FLArE and 300~MeV for FASER$\nu$2. According to Refs.~\cite{Batell:2021blf, Kling:2022ykt}, after applying these kinematic cuts, we expect less than $\mathcal{O}(1)$ neutrino-electron scattering events in the SM. Considering statistical uncertainties only, we do a log-likelihood ratio analysis where the number of expected events follow a Poisson distribution, allowing us to set limits on neutrino magnetic moment and millicharge.
Systematic uncertainties are expected to be under control since the neutrino-electron cross section is well understood and the neutrino fluxes can be constrained by the same experiment through a measurement of the event rate of neutrino charged current scattering~\cite{Kling:2023tgr}. Notably, for the considered events with small electron recoil energies, the SM event rate is small, and therefore neutrino flux uncertainties are less relevant than for neutrino charge radius measurements.

We present projected sensitivity on neutrino magnetic moment and millicharge in the upper part of \cref{tab:results}. 
The upper bounds are given for different flavors at FASER$\nu$2, FLArE, and FLArE-100, considering an integrated luminosity of 3 ab$^{-1}$ at HL-LHC. Note that the bounds are slightly sensitive to the sign of the neutrino millicharge due to the presence of the interference terms. 
\medskip 

Unlike the other two neutrino properties, the effect of a neutrino charge radius is not confined to a specific energy region. Instead, we search for an increased neutrino scattering 
event rate across the whole energy spectrum. For this, we consider both the electron scattering and nuclear scattering channel, where the latter will turn out to be more sensitive due to the significantly larger overall event rate. As this nuclear scattering channel is essentially a precision measurement of the total neutral current scattering rate, it is subject to systematic uncertainties, which we discuss below. 

One major source of systematic uncertainties is associated with the neutrino flux. While the uncertainties on the LHC neutrino flux predictions are currently large~\cite{Kling:2021gos, Bai:2020ukz, Jeong:2021vqp, Maciula:2022lzk, FASER:2024ykc}, a measurement of the charged current event rate will constrain the fluxes once the experiment starts taking data. In our analysis, we take this into account by considering the statistical uncertainty expected in the measurement of charged current events as a proxy for the uncertainty on the flux estimates.

Another source of uncertainty is associated with the modelling of the neutrino-nucleus interaction cross section. This includes, for example, parton distribution functions, quark mass effects, higher order radiative corrections, nuclear shadowing and anti-shadowing effects, the modelling of parton shower and hadronization inside the target nucleus, as well as final state interactions. A recent prediction using the NNSF$\nu$ neutrino structure functions introduced in Ref.~\cite{Candido:2023utz} quotes a 2.5 - 3\% uncertainty on the cross section for the energy range 100 GeV $<~E_{\nu}~<$ few TeV, which is relevant to this study. We will use this result as a conservative estimate. Future measurements like at the FPF~\cite{Cruz-Martinez:2023sdv} and EIC~\cite{Khalek:2021ulf, AbdulKhalek:2019mzd} will reduce these uncertainties and bring to a 1\% level. Though these uncertainties are quoted for charge current interactions, we expect them to be similar for neutral current interactions as the underlying PDFs are the same.  We illustrate the effect of this by showing results for 3\% and 1\% uncertainty. We note that cross section modelling uncertainties quoted in the weak mixing angle analysis performed by NuTeV~\cite{NuTeV:2001whx} were around 0.3\%, and therefore even smaller than the values assumed here, indicating that even a 1\% uncertainty is not unrealistic. Hence, we also show the results for 0.3\% uncertainty as an optimistic case and with 0\% uncertainty to illustrate what the best case scenario could be. 

Finally, there could be uncertainties arising from the experimental setup, for example, related to energy reconstruction, detection efficiency, particle identification, and event classification. 
In particular, very soft leptons from CC interactions can be missed and be mistaken for NC events.
Since the detector designs are still under development, the details on the detector performance are not yet available. However, this also leaves room to consider the signatures under discussion in this study as a benchmark for detector design and optimize them accordingly. In the following, we assume that detector-related uncertainties can be sufficiently reduced to be smaller than the statistical and cross section uncertainties of the measurement. 

Unlike neutrino magnetic moment and millicharge measurements, here we look at the entire energy spectrum $\left(E_{r} > E_{\rm{thr}}\right)$, and hence we expect higher event rates. So we employ a $\chi^2$ analysis to set bounds on neutrino charge radius that can be probed at these forward detectors. We consider statistical uncertainty and also systematic uncertainties coming from neutrino flux and neutrino-nucleus cross section. As mentioned above, uncertainty on the neutrino flux estimate is included by considering the statistical uncertainty in the CC events as a proxy. For systematic uncertainty on the neutrino-nucleus cross section, we include a pull parameter with an uncertainty of 3\%, 1\%, and 0.3\%.
The projected sensitivity on neutrino charge radius, considering only statistical uncertainties and systematic uncertainty coming from the neutrino flux, are presented in the lower part of \cref{tab:results}. As expected, the bounds obtained from the electron scattering signature are much weaker compared to those from nuclear scattering. In \cref{tab:NCR_systematics}, we show the results for neutrino charge radius from nuclear scattering only, including various values for cross section uncertainty. We see how going from 0\% to 3\% weakens the bound by roughly a factor of few for FLArE.
\medskip 

\begin{figure*}[th]
    \includegraphics[width=0.32\textwidth,clip]{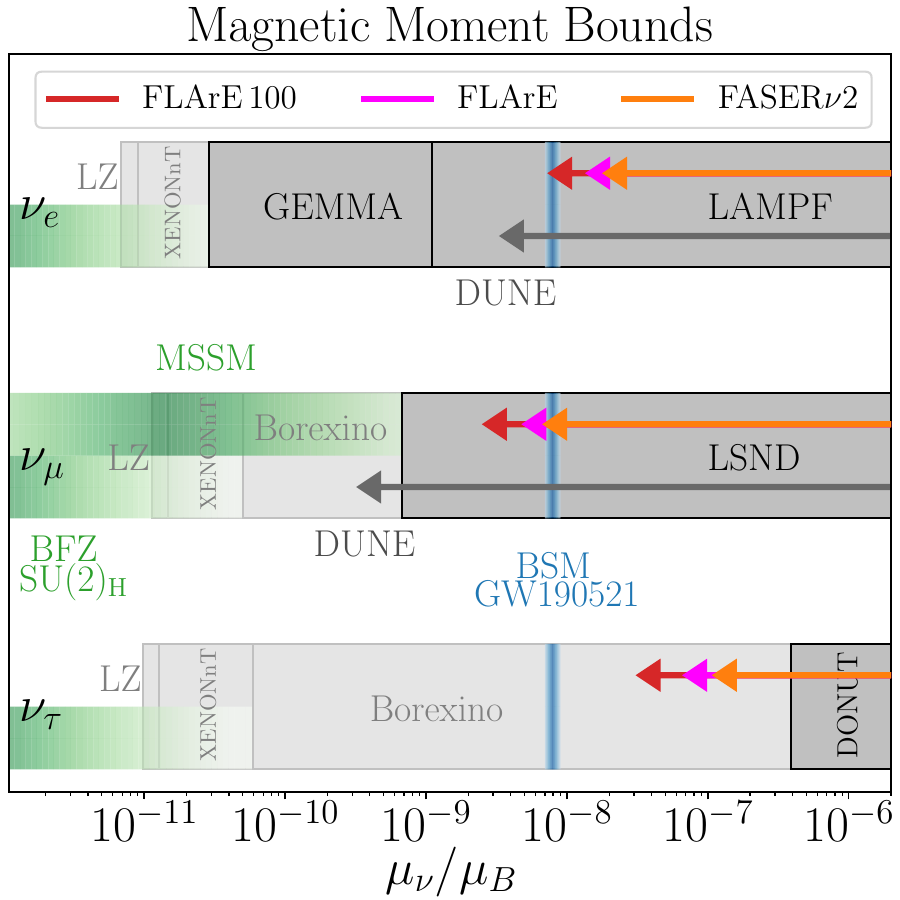} 
    \includegraphics[width=0.32\textwidth,clip]{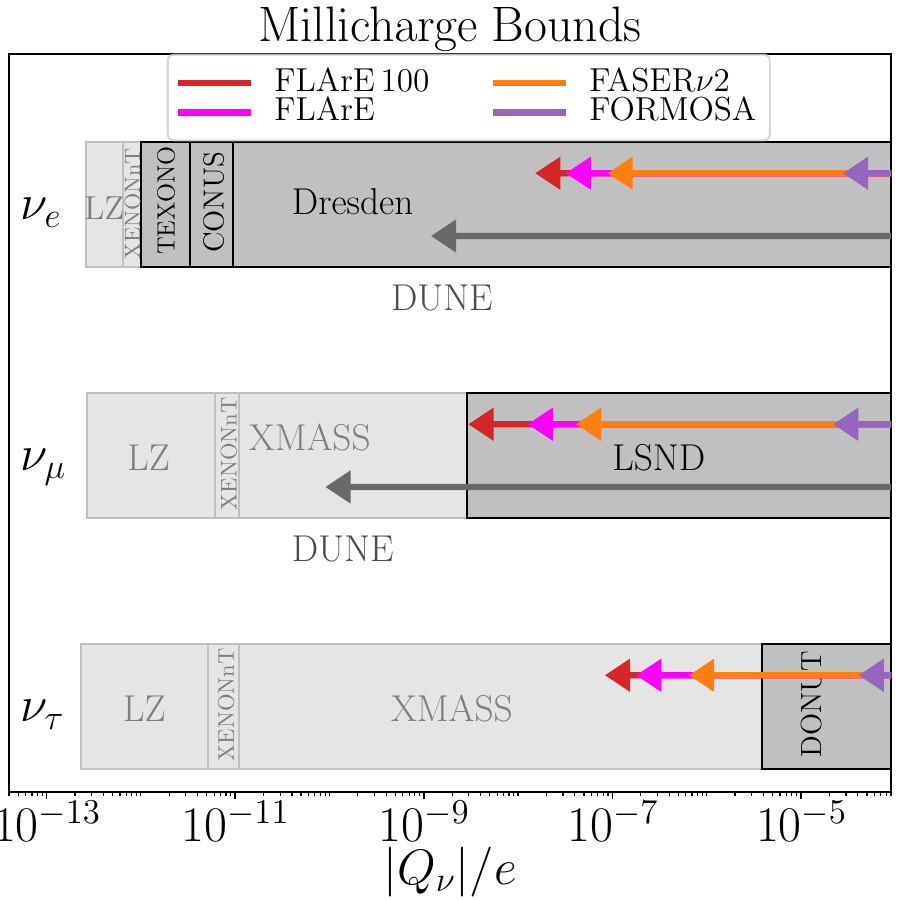}   
    \includegraphics[width=0.32\textwidth,clip]{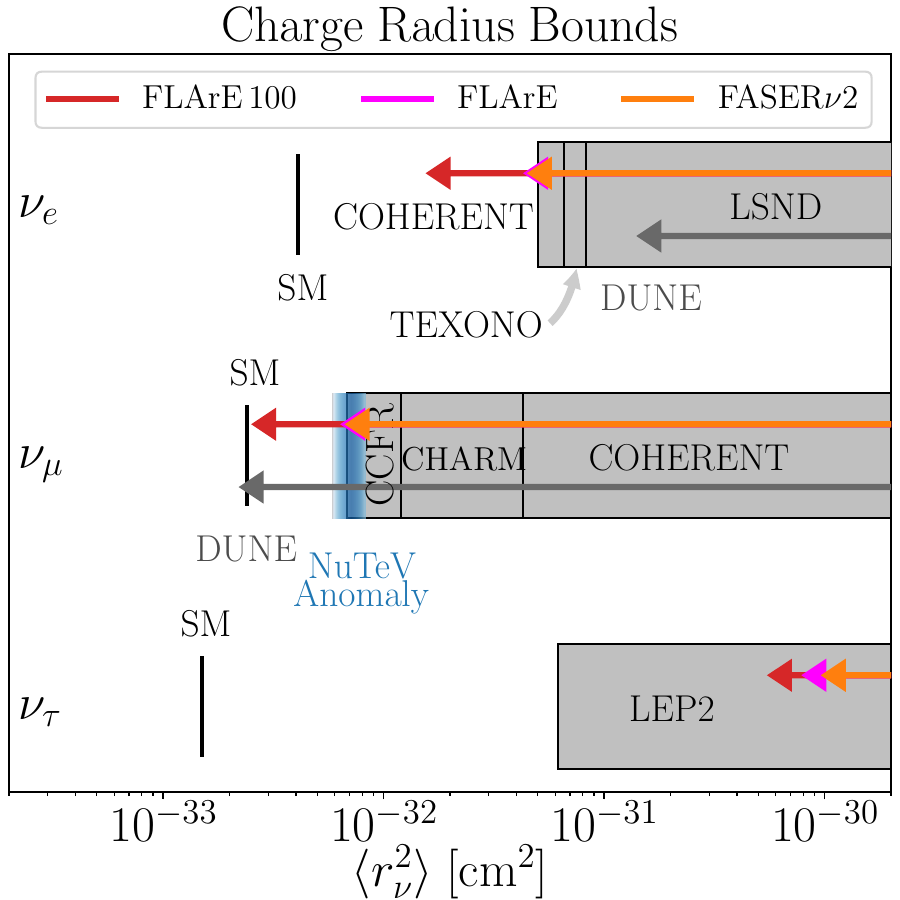} 
    \caption{Comparison of experimental bounds on neutrino electromagnetic properties: effective magnetic moment (left), millicharge (middle), and charge radius (right). The projected sensitivity of FASER$\nu$2 (orange), FLArE (magenta), and FLArE-100 (red) shown alongside existing accelerator and reactor constraints (dark gray shaded), direct detection limits from solar neutrino flux (light gray shaded) and projections from other proposed searches (gray arrow). 
    The FASER$\nu$2, and FLArE(-100) results in this plot consider only statistical and flux uncertainties.
    The blue-shaded regions correspond to the magnetic moment values that explain the NuTeV anomaly, and gravitational waves signal from black hole mergers. The contribution of BSM benchmark models to large magnetic moments is presented in green. 
    FLArE can set the world's leading limits on neutrino magnetic moment and millicharge for tau neutrino coming from terrestrially produced neutrinos, and set the world's leading limit for electron neutrino charge radius.
    The limits on muon neutrino charge radius for FLArE come within a factor of a few from the SM prediction.
    }
    \label{fig:final_barplot}
\end{figure*}

The obtained sensitives (without cross section uncertainties) to the neutrino EM properties are also presented in \cref{fig:final_barplot}, alongside existing constraints and relevant benchmark scenarios. Here we show only the positive bounds, as the negative values are very similar in absolute value. A recent projection on the sensitivity for electron and muon neutrinos at DUNE, as obtained in Ref.~\cite{Mathur:2021trm}, is also shown for comparison. 

The left panel shows the results for the neutrino magnetic moment. Shown as dark gray shaded regions are current constraints obtained by purely laboratory experiments using reactor and accelerator neutrinos from GEMMA~\cite{Beda:2012zz}, TEXONO~\cite{TEXONO:2006xds}, LAMPF~\cite{Allen:1992qe}, LSND~\cite{LSND:2001akn} and DONUT~\cite{DONUT:2001zvi}. The light gray shaded region corresponds to measurements using solar neutrinos at Borexino~\cite{Borexino:2017fbd}, XENONnT~\cite{XENON:2022ltv, Khan:2022bel, A:2022acy} and LZ~\cite{LZ:2022lsv,A:2022acy}. We can see that FLArE will be able to provide the leading sensitivity to tau neutrino magnetic moment obtained using a pure laboratory measurement and constrain $\mu_{\nu_\tau}\lesssim 7\times 10^{-8}\mu_B$. This is due to the large flux of tau neutrino at the LHC location compared with other laboratory neutrino sources.

Throughout the years, a variety of new physics models predicting large neutrino magnetic moments have been proposed~\cite{Rajpoot:1990hj, Mohapatra:2004ce, Barbieri:1988fh}. Such scenarios have been revisited recently in Refs.~\cite{Lindner:2017uvt,Babu:2020ivd}. The authors conclude that models of Dirac neutrinos with large diagonal neutrino magnetic moments do not seem possible anymore unless one is willing to accept it to be fine-tuned. An example of such a scenario was found in Ref.~\cite{Aboubrahim:2013yfa} in a scan over the MSSM parameter space, where a muon magnetic neutrino magnetic moment as large as $10^{-9}~\mu_B$ was found. In contrast, for Majorana neutrinos, large transition moments can be realized, for example using a $SU(2)_H$ horizontal symmetry~\cite{Babu:1989wn} or a BFZ model~\cite{Barr:1990um}. We illustrate those scenarios as the green region in \cref{fig:final_barplot}. 
In addition, large neutrino magnetic moments have been used to explain the existence of black holes in the mass-gap region that have been detected in the gravitational wave event GW190521~\cite{Sakstein:2020axg}. The corresponding region is marked in blue.

We present the results for neutrino millicharge in the middle panel of \cref{fig:final_barplot}. As before, the dark-shaded regions show purely laboratory constraints from DONUT~\cite{DONUT:2001zvi,Das:2020egb}, LSND~\cite{LSND:2001akn,Das:2020egb}, Dresden-II~\cite{AtzoriCorona:2022qrf}, CONUS~\cite{CONUS:2022qbb}, TEXONO~\cite{Gninenko:2006fi,Chen:2014dsa}, and GEMMA~\cite{Studenikin:2013my,Chen:2014dsa}. Upper limits on neutrino millicharge have been also obtained using solar neutrinos by XMASS~\cite{XMASS:2020zke}, XENONnT~\cite{XENON:2022ltv, Khan:2022bel, A:2022acy} and LZ~\cite{LZ:2022lsv,A:2022acy} as shown by the light gray shaded regions. Not included in this figure are additional constraints from astrophysical considerations, since they are subject to additional underlying assumptions and uncertainties compared to pure laboratory constraints. In particular, the neutrino millicharge can have an impact on astrophysical phenomena such as red giant or solar cooling~\cite{Raffelt:1999gv}, the rotation of magnetized neutron stars~\cite{Studenikin:2012vi}, and the arrival time of SN 1987A supernova neutrinos~\cite{Barbiellini:1987zz}, resulting in approximate upper limits on the effective charge of electron neutrino in the range $|Q_{\nu_e}|\lesssim 10^{-14}-10^{-19}$. Even stronger constraints than astrophysical arguments on the electron neutrino millicharge can be obtained from electric charge conservation in neutron beta decay, along with the experimental bounds on the neutron charge and the non-neutrality of matter giving $\abs{Q_{\nu_e}}\lesssim 10^{-21}e$~\cite{Raffelt:1999gv}. 
We find that FLArE is potentially capable of providing the most stringent limit on the effective electric charge of tau neutrino from terrestrially produced neutrinos, with an upper limit of $\abs{Q_{\nu_\tau}}\lesssim 10^{-7}e$. 

Following the study of millicharged particles using the scintillator-based experiment~\cite{Kelly:2018brz,Foroughi-Abari:2020qar}, we can expect to bound the neutrino millicharge at FORMOSA to $|Q_{\nu_e}|\lesssim 2.8\times10^{-5}e$, $|Q_{\nu_\mu}|\lesssim 2.2\times10^{-5}e$, and $|Q_{\nu_\tau}|\lesssim 4.1\times10^{-5}e$ with $90\%$ C.L., corresponding to a scintillator detector with quadruple coincidence. These upper bounds on neutrino millicharge, which are weaker than FLArE results, are presented in the middle panel of \cref{fig:final_barplot}. These projected sensitivities are almost independent of the neutrino flux, as the sensitivity is limited by the fact that below $Q_\nu \sim 5\times10^{-4}e$, the probability of photoelectron production drops significantly. The analysis of FORMOSA with 4 layers is considered almost background free. To demonstrate the sensitivity reach, we also assume zero background for a detector with triple coincidence and find the $90\%$ C.L. upper bounds $|Q_{\nu_e}|\lesssim 0.8\times10^{-5}e$, $|Q_{\nu_\mu}|\lesssim 0.5\times10^{-5}e$, and $|Q_{\nu_\tau}|\lesssim 1.3\times10^{-5}e$. This background-free assumption can in principle be achieved, for example, by using better photomultiplier tubes (PMTs) that have less background noise.

The right panel shows the results for the neutrino charge radius. The dark gray shaded regions are current constraints obtained by purely laboratory experiments using reactor and accelerator neutrinos from COHERENT~\cite{AtzoriCorona:2022qrf,Khan:2022akj}, CHARM-II~\cite{CHARM-II:1994aeb}, LSND~\cite{LSND:2001akn}, CCFR~\cite{CCFR:1997zzq,Hirsch:2002uv}, LEP2~\cite{Hirsch:2002uv}, TEXONO~\cite{TEXONO:2009knm}. FLArE can set the world's leading limit for electron neutrino and set highly competitive limits for muon neutrino where it comes within a factor of a few from the SM prediction. The deviation of the weak mixing angle from the SM observed by the NuTeV Collaboration~\cite{NuTeV:2001whx} could also be interpreted as a measurement of the muon neutrino charge radius $\langle r^2_{\nu_\mu} \rangle = 4.20 \times 10^{-33}$ within $1\sigma$ error~\cite{Hirsch:2002uv}. The $1 \sigma$ preferred region to explain the NuTeV anomaly is shown by the blue target region.\footnote{While there exist many possible ways to explain this anomaly with new physics~\cite{Davidson:2001ji}, a reassessment of the NuTeV results~\cite{Brodsky:2004qa,Bentz:2009yy} with more careful considerations suggest that this result is in agreement with the SM prediction.} For comparison, DUNE is expected to constrain $|\langle r^2_{\nu_\mu} \rangle| < 2\times 10^{-33}$ cm$^2$ and $|\langle r^2_{\nu_e} \rangle| < 1\times 10^{-31}$ cm$^2$, which is an order of magnitude weaker than the FLArE bound for the electron neutrino. The DUNE projection considered the electron scattering signature, which suffers from lower event rates. A measurement using nuclear scattering at DUNE does not seem promising due to the large nuclear uncertainties in the cross section for GeV energy neutrinos. 

\section{Measurement of the Weak Mixing Angle and the NuTeV Anomaly}\label{WeakMixingAngle}

\begin{figure*}[th]
    \includegraphics[width=0.8\textwidth]{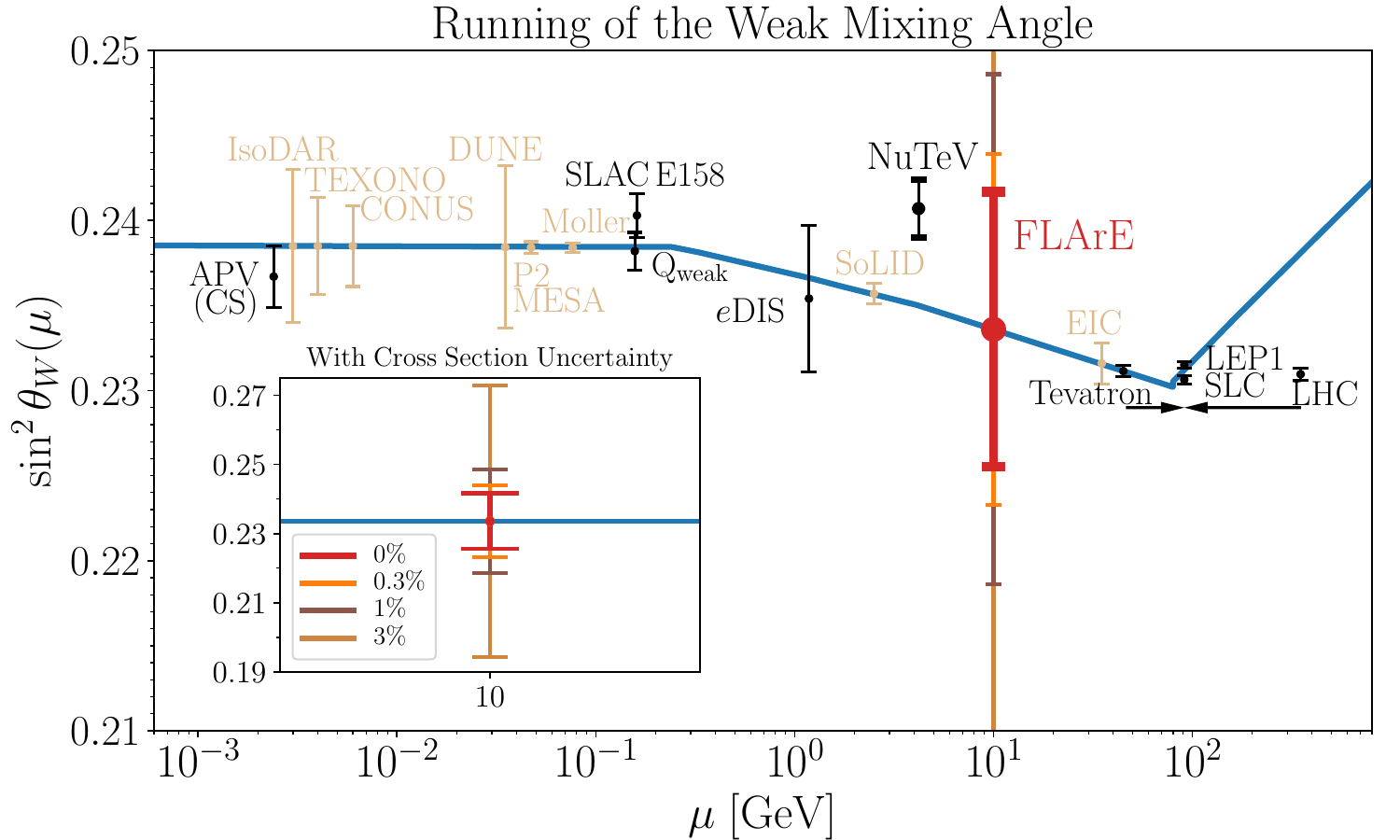}  
    \caption{Scale dependence of the weak mixing angle in the $\overline{\rm{MS}}$ scheme, $\sin^2\theta_W(\mu)$, shown with the existing measurements (black), the expected sensitivity of FLArE at FPF (red), and other future experiments (beige). The blue line corresponds to the SM prediction for the running of $\sin^2\theta_W$ with scale $\mu$. For clarity, the Tevatron and LHC points are shifted horizontally on either side.
    \textbf{Inset:} We also show the sensitivity only at FLArE including a 3\%, 1\%, 0.3\% and 0\% systematic uncertainty on the neutrino-nucleus cross section.
    }
    \label{fig:weinberg}
\end{figure*}
The measurement of neutrino interactions at the energies accessible at the FPF provides an opportunity to measure precisely the electroweak parameters. The weak mixing angle, $\sin^2\theta_W$, is one key parameter that parametrizes sseveral measurable observables in the electroweak sector of the SM. The value of $\sin^2\theta_W$ gets radiative corrections and depends on the renormalization prescription~\cite{Erler:2004in}, where $\overline{\rm{MS}}$ (modified minimal subtraction) scheme is conventionally employed. One of the best measurements of the weak mixing angle comes from Z-pole observables~\cite{ALEPH:2010aa,Workman:2022ynf} (Tevatron, LEP1, SLC, LHC) with an average value of $\sin^2\theta_W(m_{Z})_{\overline{\rm{MS}}}=0.23125(16)$, comparable to the SM value of $\sin^2\theta_W(m_{Z})_{\overline{\rm{MS}}}=0.23122(4)$~\cite{Workman:2022ynf}. At relatively low energy scales, several experimental measurements of weak mixing angle exist (for a review, see Ref.~\cite{Kumar:2013yoa}) including the electron-deep inelastic scattering~\cite{PVDIS:2014cmd} (eDIS), neutrino-nucleus scattering~\cite{NuTeV:2001whx} (NuTeV), atomic parity violation~\cite{Wood:1997zq,Guena:2004sq,Dzuba:2012kx,Safronova:2017xyt} (APV on cesium), Moller scattering~\cite{SLACE158:2005uay} (SLAC E158), elastic electron-proton scattering~\cite{Qweak:2018tjf} ($Q_{\rm{weak}}$), and coherent elastic neutrino-nucleus scattering at COHERENT~\cite{COHERENT:2021xmm} and Dresden-II~\cite{Majumdar:2022nby}. 
The precise measurement of the weak mixing angle at different energy scales provides a direct probe of new physics beyond the  SM~\cite{Kumar:2013yoa,Davoudiasl:2012qa,Davoudiasl:2015bua}. In particular, it will allow one to test the NuTeV anomaly~\cite{NuTeV:2001whx}. Using neutrino scattering, the collaboration measured a value of $\sin^{2}\theta_W$ that was 3$\sigma$ above the SM prediction at a scale of $\sim 4~\gev$. A measurement at FPF will shed more light on the running of the weak mixing angle at a similar energy scale. Any change in the weak mixing angle from the SM value, $\sin^2 \theta_W \to \sin^2 \theta_W + \Delta \sin^2 \theta_W$, will result in a shift in the vector coupling constant, 
\begin{equation}\label{VcouplingshiftfromCRforquarks}
g^{q}_{V}\rightarrow g^{q}_{V} - 2Q_q \Delta \sin^{2}\theta_W.
\end{equation}
The phenomenological consequences of this shift are therefore very similar to the study of neutrino charge radius presented in the previous section. We perform a similar analysis to obtain the FLArE expected sensitivity to the weak mixing angle and constrain $\Delta \sin^{2}\theta_W < 0.0077$ at 68$\%$ C.L. The estimate for sensitivity to $\sin^2\theta_W$ in the $\overline{\rm{MS}}$ scheme at the scale $\mu \sim Q \sim 10~\gev$, which is the typical momentum transfer for a TeV scale energy neutrino at FLArE, is shown in \cref{fig:weinberg}, along with the existing constraints and the running of the coupling predicted by the SM~\cite{Erler:2004in, Erler:2017knj, Workman:2022ynf}. Results are shown for different values of cross section uncertainty.
For comparison, we also show the projected sensitivities to the weak mixing angle from future experiments including DUNE~\cite{deGouvea:2019wav}, EIC~\cite{Boughezal:2022pmb}, Moller at JLAB~\cite{SLACE158:2005uay,Mammei:2012ph}, MESA-P2~\cite{Berger:2015aaa}, SoLID at JLAB~\cite{Souder:2012zz}, IsoDAR at Yemilab~\cite{Alonso:2021kyu}, and reactors~\cite{Kosmas:2015vsa, Canas:2016vxp, Canas:2018rng, Lindner:2016wff} (TEXONO, CONUS).

A precise measurement of the weak mixing angle requires good control over various systematic uncertainties. These are analogous to the measurement of the neutrino charge radius through nuclear scattering, and we refer the reader to the previous section for a more detailed discussion. As before, we have considered statistical uncertainties as well as uncertainties associated with flux normalization as constrained by charged current measurements, and neutrino-nucleus cross section in our sensitivity estimate. It is worth noting that the measurement of the weak mixing angle provides a well-motivated benchmark for detector performance requirements that should be considered during detector design.

\section{Conclusion}\label{conclusion}

The immense flux of neutrinos in the forward region of the LHC provides an excellent opportunity for neutrino physics. This neutrino beam is a powerful source of the most energetic human-made neutrinos for all three flavors. The proposed neutrino detectors at the FPF, FASER$\nu$2 and FLArE, can use this neutrino beam to set stringent constraints on neutrino electromagnetic properties and measure the weak mixing angle to percent level precision.

In this paper, we have presented a detailed phenomenological study on the potential of the FPF experiments to probe the neutrino electromagnetic properties: magnetic moment, millicharge and charge radius. All these scenarios result in an excess of neutral current events that can be observed at these detectors. We first look at neutrino-electron elastic scattering, where in the presence of neutrino magnetic moment and millicharge the excess events are at low electron recoil energies. 
Focusing on this kinematic region and taking advantage of the huge tau neutrino flux, FPF can set the strongest limits on neutrino magnetic moment and millicharge for tau neutrinos coming from terrestrially produced neutrinos.
For neutrino charge radius, better constraints are obtained by looking at the neutral current neutrino DIS process, where the heavier target results in an increased event rate over neutrino-electron elastic scattering. By looking for excess events across the entire spectrum, FPF can set the world's leading limits on the neutrino charge radius for electron neutrinos and, for muon neutrinos, FPF can come within a factor of a few from the SM prediction. We have summarized our results in \cref{tab:results} and \cref{fig:final_barplot}.

An important test of the SM is the measurement of electroweak parameters at different energy scales. FPF has the potential to measure the weak mixing angle with a precision of about $3\%$ at an energy scale of $\mu \sim 10~\gev$. In \cref{fig:weinberg}, we show the scale dependence of the weak mixing angle along with the FPF measurement, which considers both statistical and flux uncertainties. This is an important test of the SM, especially in light of the NuTeV anomaly. The ability to measure the weak mixing angle with high precision sets an important benchmark for the design of the FPF neutrino detectors.

\section{Acknowledgment}

We thank Adam Ritz, Zahra Tabrizi, and Sebastian Trojanowski for fruitful discussions. We are grateful to the authors and maintainers of many open-source software packages, including FORESEE~\cite{Kling:2021fwx}, and LHAPDF~\cite{Buckley:2014ana}. F.K. acknowledges support by the Deutsche Forschungsgemeinschaft under Germany's Excellence Strategy - EXC 2121 Quantum Universe - 390833306.
R.M.A. is supported in part by the DOE under Grant No. DE- SC0016013. 
S.F. is supported in part by NSERC, Canada. 
Y.-D.T. is supported in part by NSF Grant No.~PHY-2210283 and in part by Simons Foundation Grant No.~623683. Y.-D.T also thanks the support from the LANL Director's Fellowship. This research is partially supported by LANL's Laboratory Directed Research and Development (LDRD) program. 

\bibliography{references}

\end{document}